\documentclass[proof]{WileyASNA-v1}
\articletype{Article Type}%
\received{20 May 2020}
\revised{}
\accepted{}
\begin{document}
\title{Properties of white dwarf in the binary system AR Scorpii and its observed features}

\author[1,2]{K. K. Singh*}

\author[2]{P. J. Meintjes}

\author[1,3]{K. K. Yadav}

\authormark{K. K. Singh, P. J. Meintjes \& K. K. Yadav}

\address[1]{\orgdiv{Astrophysical Sciences Division}, \orgname{Bhabha Atomic Research Centre}, \orgaddress{\state{Mumbai 400085}, \country{India}}}

\address[2]{\orgdiv{Physics Department}, \orgname{University of the Free State}, \orgaddress{\state{Bloemfontein 9300}, \country{South Africa}}}

\address[3]{\orgname {Homi Bhabha National Institute}, \orgaddress{\state{Mumbai 400094}, \country{India}}}

\corres{*K. K. Singh, \email{kksastro@barc.gov.in}}

\presentaddress{Astrophysical Sciences Division, Bhabha Atomic Research Centre, Mumbai 400085, India}

\abstract{The binary system AR Scorpii hosts an M-type main sequence cool star orbiting around a magnetic white dwarf in 
the Milky Way Galaxy. The broadband non-thermal emission over radio, optical and X-ray wavebands observed from AR Scorpii 
indicates strong modulations on the spin frequency of the white dwarf as well as the spin-orbit beat frequency of the system. 
Therefore, AR Scorpii is also referred to as a white dwarf pulsar wherein a fast spinning white dwarf star plays very crucial 
role in the broadband non-thermal emission. Several interpretations for the observed features of AR Scorpii appear in 
the literature without firm conclusions. In this work, we investigate connection between some of the important physical properties 
like spin-down power, surface magnetic field, equation of state, temperature and gravity associated with the white dwarf in the binary 
system AR Scorpii and its observational characteristics. We explore the plausible effects of white dwarf surface magentic field on 
the absence of substantial accretion in this binary system and also discuss the gravitational wave emission due to 
magnetic deformation mechanism.}

\keywords{binaries: white dwarfs, individual: AR Scorpii, physical-properties: general}

\maketitle


\section{Introduction}\label{sec1}
White dwarfs (WDs) belong to the family of relativistic compact objects in the Universe. WDs along with other compact objects 
like neutron star and black hole have extreme physical conditions which cannot be produced in a terrestrial laboratory. 
Hence, they provide natural sites to explore several fundamental laws of physics \citep{Barack2019}. Classically, the astrophysical 
compact objects are formed through the gravitational collapse of a normal star which has run out of its thermonuclear fuel for fusion. 
The core of such exhausted star is no longer able to produce sufficient radiation pressure to prevent its outer layer from 
gravitational collapse. In Newtonian gravitation, the mass of collapsing star ($M$) is constant even if it radiates \citep{Mitra2013}. 
Therefore, as radius of star ($R$) decreases during the collapse, the ratio of mass to radius increases and leads to the extreme density in 
the final stable stage of the stellar evolution. This is characterized by the dimensionless compactness parameter defined as  
\begin{equation}\label{compactness}
	\eta~=~\frac{2GM}{c^2 R}~=~0.3\left(\frac{M}{M_\odot}\right)\left(\frac{\rm 10^6~cm}{R}\right)
\end{equation}
where $G$ is the universal gravitational constant, $M_\odot$ is the solar mass and $c$ is the speed of light in vacuum. 
In the general theory of relativity, energy and momentum sources associated with a massive object contribute to its gravity. 
Therefore, the energy density or pressure not only opposes the self gravity of the star but increases it also. 
The enhancement of gravity due to pressure becomes dominant with increasing value of $\eta$ and sets a critical mass beyond 
which gravitational collapse cannot be avoided. However, the collapse of WDs is prevented by the hydrostatic equilibrium between 
the inward gravity and outward pressure of the relativistic degenerate electron gas \citep{Chandrasekhar1931a,Chandrasekhar1931b}. 
Static or non-rotating ideal WDs have a mass ($M_{wd}$) less than a critical value given by the Chandrasekhar 
limit ($M_{ch}=1.44 M_\odot$) and a radius ($R_{wd}$) same as that of the Earth ($\sim 10^4$ km). If mass of the static 
WD exceeds the Chandrasekhar limit ($M_{wd} > M_{ch}$), it becomes unstable and yields to type Ia supernova (SNe Ia) explosion which 
is an excellent distance indicator in the cosmological observations. In the presence of strong magnetic fields or fast rotations, 
the mass of a WD can exceed the Chandrasekhar limit \citep{Das2014,Ostriker1968}. The progenitors of WDs have very low mass where 
thermonuclear reactions are inhibited before synthesis of iron (Fe). WDs are mainly classified as Helium (He-WD), Carbon-Oxygen (CO-WD) 
and Oxygen-Neon (ONe-WD) with the increasing progenitor or initial stellar masses and their underlying composition \citep{Camisassa2019}. 
Because of the high surface gravitational field due to a large compactness of WDs, heavy elements are diffused downward and therefore 
the atmosphere around WDs contain mainly light elements like hydrogen (H) and He. From the presence of spectral lines, WDs are divided 
into two groups namely DA and DB \citep{Kleinman2013}. WDs belonging to DA group have H-rich atmosphere and represent approximately 
80$\%$ of the population. The atmosphere of remaining 20$\%$ DB group WDs is dominated by He. It is believed that CO-WDs in binaries 
form SNe Ia when their masses reach the Chandrasekhar limit \citep{Wang2012}.
\par
WDs account for the final fate of more than 95$\%$ of all the stellar population in the Milky Way and are therefore considered as 
the most most abundant stellar remnants \citep{Parsons2020}. The average mass and density of WDs are 0.6$M_\odot$ and 10$^6$ g~cm$^{-3}$ 
respectively. A significant fraction of the WD population has a very high magnetic field with strengths in the range 10$^3$ G-10$^9$ G. 
Isolated rotating WDs and WDs in binary systems are prominent Galactic sources in the Universe. In this work, we investigate the 
physical properties of the WD parameters in the binary system AR Scorpii which has been discovered as a pulsating emitter at wavelengths 
from radio to X-ray \citep{Marsh2016}. The paper is organized as the following. A brief description of the observed features of the binary 
system AR Scorpii is given in Section \ref{sec2}. In Section \ref{sec3}, we derive and discuss various parameters from the 
physical properties of WD in AR Scorpii. Finally, we have summarized the study in Section \ref{sec4}.
\section{An overview of AR Scorpii}\label{sec2}
AR Scorpii was discovered as a binary system located in the ecliptic plane near to the Galactic centre at a distance of 
d $\sim$ 115 pc from the Earth \citep{Marsh2016}. It consists of a fast spinning magnetized WD and an M-type main sequence 
cool star (also referred to as M-dwarf or red-dwarf) as companion or donor. Both revolve around common centre-of-mass with an 
orbital period of $P_o =$ 3.56 hours. The binary separation between WD and companion star is $ a \sim 7.6\times10^{10}$ cm. 
A geometrical model suggests that the magnetic WD in AR Scorpii is a perpendicular rotator and both open field lines periodically 
sweep the stellar wind of the companion star \citep{Geng2016}. It means the spin axis of WD is orthogonal to the magnetic axis.
A bow shock is formed through the interaction between the stellar wind of donor star and beam of particles streaming out from 
the open magnetic field line regions. This shock propagates into the stellar plasma and accelerates the electrons in the wind 
to relativistic energies. The accelerated electrons produce synchrotron radiation extending from radio/optical to X-ray 
wavelengths \citep{Geng2016}. Broadband non-thermal emission observed from AR Scorpii is modulated over the spin frequency of WD and 
spin-orbit beat frequency of the binary system \citep{Marsh2016,Buckley2017,Takata2018,Stanway2018,Garnavich2019,Singh2020}. 
Therefore, the binary system AR Scorpii is also dubbed as a WD pulsar which involves interaction between the magnetospheres of two 
stars. However, the observed unique features of AR Scorpii are very different from the normal radio pulsars.


\section{Properties of white dwarf in AR Scorpii}\label{sec3}
The pulsed emission from the binary system AR Scorpii is assumed to be powered by a rapidly spinning magnetic WD through the ultimate 
conversion of rotational energy into radiation. From the observations of AR Scorpii, the mass and radius of WD have been 
estimated as $M_{wd}= 0.8M_\odot$ and $R_{wd}= 7\times10^3$ km respectively \citep{Marsh2016}. The spin period and its time-derivative 
for WD are originally derived as $P_s = 117$ s (1.95 minutes) and $\dot P_s = 3.9\times10^{-13}$ s s$^{-1}$ respectively. A more extensive 
observation involving high time resolution photometry later \citep{Stiller2018} suggests $\dot P_s = 7.18\times10^{-13}$ s s$^{-1}$ which 
is about twice the original value derived by \citet{Marsh2016}. From these observed parameters, we investigate various physical properties 
of WD in AR Scorpii under following Subsections.

\subsection{Mechanical}
The mass of WD ($M_{wd}$) and ratio ($q$) of companion star mass to the WD mass are important initial parameters to describe a 
binary system. For AR Scorpii the mass ratio is $q=$0.375. The mass ratio distribution helps in exploring the evolution of companion 
stars over a given mass range and estimation of the binary separation ($a$). For $q=$0.375, we get \citep{Buckley2017} 
\begin{equation}
	a~\approx~7.6\times10^{10} \left(\frac{M_{wd}}{0.8 \rm M_\odot}\right)^{\frac{1}{3}} \left(\frac{P_o}{1.28\times10^4~ \rm s}\right)^{\frac{2}{3}}~~\rm cm
\end{equation}
From the mechanics of uniform solid sphere, the moment of inertia of WD is given by  
\begin{equation}
		I_{wd}~=~\frac{2}{5}M_{wd}R^2_{wd}
\end{equation}
and the corresponding rotational energy of the spinning WD is  
\begin{equation}
	E_{rot}~=~\frac{1}{2}I_{wd}\Omega^2_{s}~=~2\pi^2I_{wd}P_s^{-2}
\end{equation}
where $\Omega_s =2\pi P^{-1}_s$ is the spin angular frequency of WD. For WD in AR Scorpii, $I_{wd} = 3.1\times10^{50}$ g~cm$^2$ and 
$E_{rot} = 4.5\times10^{47}$ erg. This rotational energy is released as the spin-down power or luminosity ($L_{sd}$) of the system, 
which can be expressed as 
\begin{equation}
	L_{sd}~=~-\frac{d}{dt}\left(E_{rot}\right)~=~4\pi^2I_{wd}\dot P_sP^{-3}_s
\end{equation}
From the measured parameters of WD in the binary system AR Scorpii, we get $L_{sd} = 3.0\times10^{33}$ erg~s$^{-1}$. 
The mean luminosity of AR Scorpii for non-thermal emission is $L_{non-th} \sim 1.6\times10^{31}$ erg~s$^{-1}$ \citep{Marsh2016}. 
\citet{Takata2018} have estimated the X-ray luminosity of AR Scorpii as $L_X \sim 4.0\times10^{30}$ erg~s$^{-1}$ from the 
observations. This implies that both $L_{non-th}$ and  $L_X$ are even less than 1$\%$ of the spin-down luminosity of WD, 
suggesting that the rotational energy loss of the spinning WD can sufficiently power the non-thermal emission from the binary 
system AR Scorpii. This is similar to the properties of radiation emitted from the spin-powered neutron star radio pulsars. 
The braking timescale over which rotational energy is released or the spin-down time of WD is given by 
\begin{equation}
	\tau_{sd}~=~\frac{P_s}{\dot P_s}~\sim~\frac{E_{rot}}{L_{sd}}
\end{equation}	
From the above estimates, the value of $\tau_{sd}$ suggests that the extraction of rotational energy of WD in the 
binary system AR Scorpii will slow down its spin on a timescale of $\sim 10^7$ years. 

\subsection{Thermodynamical}
A complete thermodynamical characterization of matter under extreme physical conditions is given by the equation of state 
which describes the pressure as a function of density and temperature. The average density of a compact object is defined as 
\begin{equation}
	\rho~=~\frac{3M}{4\pi R^3}~=~~4.72\times10^{14}\left(\frac{M}{M_\odot}\right)\left(\frac{\rm 10^6~cm}{R}\right)^3~~ \rm g~cm^{-3}
\end{equation}	
For WD in AR Scorpii, we get $\rho_{wd} =1.1\times10^6$ g cm$^{-3}$. A WD with such a high density is in hydrostatic 
equilibrium due to balance between the gravitational contraction and electron degeneracy pressure. The degeneracy pressure 
of high density electron gas in WD results from the Pauli exclusion principle in quantum statistics which states that 
all the fermions (electrons, protons, neutrons, etc) cannot occupy the lowest energy level  in a gravitationally bound system 
and therefore high energy levels are occupied. The pressure of degenerate electron gas ($P_{deg}$) is independent of 
temperature and the thermal pressure ($P_{th}$) of a WD is negligible. The equation of state of matter in WD is modeled 
using different approaches, the method proposed by \citet{Chandrasekhar1931a} is widely used to describe the properties of 
WDs. In 1926, Fowler assumed non-relativistic motion of electrons inside a dense matter distribution like core of WDs and 
showed that \citep{Fowler1926}
\begin{equation}
	P_{deg}~\propto~\rho^{5/3}
\end{equation}	
In 1931, Chandrasekhar used relativistic Fermi-Dirac statistics for degenerate electron gas and demonstrated 
that \citep{Chandrasekhar1931a} 
\begin{equation}
	P_{deg}~\propto~\rho^{4/3}
\end{equation}
 In this approach, the total energy density inside a WD is determined by $P_{deg}$ and energy density of nuclei without 
 considering atmosphere of WD. Using this equation of state, the degeneracy pressure of a WD for a finite temperature of 
 the core is given by \citep{Chandrasekhar1931a}
\begin{equation}
	P_{wd}~=~3.1\times10^{14}~\rho_{wd}^{4/3}~~ \rm erg~cm^{-3}
\end{equation}
For WD in AR Scorpii, the electron degeneracy pressure is estimated as $P_{wd} \sim 4\times10^{20}$ erg cm$^{-3}$. 
A surface temperature of WD in the range of 9750 K - 12000 K has been measured from various observations of 
the binary system AR Scorpii \citep{Marsh2016}. From the Koester formula, the relation between effective surface 
temperature ($T_{wd}$) and core temperature ($T_c$) of a WD is given by \citep{Koester1976}
\begin{equation}
	T_{wd}~=~3.8\times10^{-3}~g_{wd}^{0.25}~T_c^{0.64}~~\rm K
\end{equation}
where $g_{wd}$ is the surface gravity of WD. For $T_{wd} =$ 9750 K as estimated by \citet{Marsh2016}, we get $T_c \sim 10^7$ K. 
This indicates that the core of WD in the binary system AR Scorpii is about three orders of magnitude hotter than its surface. 
The thermal luminosity from the surface of a WD can be expressed as 
\begin{equation}
	L_{th}~=~4\pi \sigma_{SB} R_{wd}^2 T_{wd}^4~=~7.0 \times 10^{32}\left(\frac{R_{wd}}{10^6 \rm cm}\right)^2
	         \left(\frac{T_{wd}}{10^6 \rm K}\right)^4~\rm erg~s^{-1} 
\end{equation}
where $\sigma_{SB}$ is the Stefan-Boltzmann constant. For the case of WD in AR Scorpii, the thermal luminosity is estimated as 
$L_{th} \approx 3.5\times10^{30}$ erg s$^{-1}$ which is about three orders of magnitude less than the spin-down luminosity. 
This implies that the non-thermal emission powered by the fast spinning of WD dominates over the thermal emission of WD in 
AR Scorpii binary system. The thermal age of spinning WD from the surface temperature 
\begin{equation}
	\tau_{th}~\approx~\frac{E_{rot}}{L_{th}}
\end{equation}
gives $\tau_{th} \sim 10^9$ years which is much larger than spin-down timescale $\tau_{sd}$ 
of WD in AR Scorpii. This suggests that WD in AR Scorpii shines at cost of cooling over cosmic time longer than 
its spin slowing down timescale.

\subsection{Gravitation}
The compactness parameter ($\eta$) plays an important role to describe the gravitational properties of the compact objects. 
The strength of gravity of a compact object is measured in terms of $\eta$. From Equation \ref{compactness}, the value of compactness 
parameter for WD in AR Scorpii is estimated as $\eta_{wd} = 3\times10^{-4}$, which is two orders of magnitude larger than that of the Sun.
This indicates that the binding energy, which is directly proportional to the compactness parameter, of WD in the binary system is also very 
high as compared to the Sun. The escape velocity from the surface of a compact object or WD is defined as 
\begin{equation}
	v_{esc}~=~\sqrt{\frac{2GM}{R}}~=~3\times10^{10}~\sqrt{\eta}~~ \rm cm~s^{-1}
\end{equation}
For $\eta = 3\times10^{-4}$, we get the escape velocity for WD in AR Scorpii as $\sim$ 5000 km~s$^{-1}$, which is approximately 500 times the 
escape velocity on the Earth-surface. This implies that the gravitational field of  WD in AR Scorpii is much stronger than that of the Earth.
The surface gravity of WD estimated as 
\begin{equation}
	g_{wd}~=~\frac{GM_{wd}}{R_{wd}^2}~=~2.2\times10^8~\rm cm~s^{-2}
\end{equation}	
is about six order of magnitude larger than that of the Earth. From the general theory of relativity, the space-time curvature around a 
massive body is also defined to characterize the strength of gravity. The space-time curvature at the surface of a compact object
can be expressed as 
\begin{equation}
	\kappa~=~\frac{4\sqrt{3}GM}{R^3 c^2}~=~10^{-12} \left(\frac{M}{M_\odot}\right)\left(\frac{10^6~\rm cm}{R}\right)^3~~\rm cm^{-2}
\end{equation}	
For WD in AR Scorpii, we obtain $\kappa =2.3\times10^{-21}$ cm$^{-2}$ which is about five order of magnitude higher than the space-time 
curvature at the surface of Earth. Thus, the surface gravity and space-time curvature parameters estimated for WD in AR Scorpii suggest 
that the strength of gravity associated with the primary star in the binary system is much stronger than that of the Earth. Therefore, 
it is expected that the WD in binary system AR Scorpii should accrete matter from the stellar wind of the companion M-star through the 
gravitational capture. If the stellar plasma supplied by companion star has a large angular momentum due to its orbital motion, an accretion 
disk should be ultimately formed around WD in the binary plane of AR Scorpii. The luminosity produced by the accretion process of the WD is 
given by
\begin{equation}
	L_{acc}~=~\frac{GM_{wd}\dot m}{R_{wd}}
\end{equation}
where $\dot m$ is the accretion rate. The amount of gravitational energy release per unit mass is 
\begin{equation}
	\Delta E_{acc}~=~\frac{GM_{wd}}{R_{wd}}=~1.32\times10^{20}\left(\frac{M}{M_\odot}\right)\left(\frac{10^6~\rm cm}{R}\right)~\rm erg~g^{-1}
\end{equation}	
For WD in AR Scorpii, $\Delta E_{acc} =1.50\times10^{17}$ erg g$^{-1}$. 
The gravitational potential energy released on reaching the stellar surface provides heat for converting the accreted energy into electromagnetic 
radiation through basic process of thermal emission from the heated WD surface and hot plasma around it. Assuming that the observed X-ray luminosity is 
powered by the accretion from the companion star i.e $L_X = L_{acc}$, we get $\dot m \sim 2.6\times10^{13}$ g~s$^{-1}$, which is about two order of 
magnitude lower than the accretion rate for typical white dwarfs ($\dot m \sim 10^{15}$ g~s$^{-1}$). Thus, the little X-ray luminosity ($\sim 4\%$ of the optical luminosity) 
measured from AR Scorpii \citep{Marsh2016} together with the absence of Doppler broadened spectral emission lines suggest that very small accretion power 
is produced in this binary system. Moreover, the high spin-down rate of WD observed today implies a previous high accretion rate responsible for its spin up stage, 
which contradicts the lack of observational evidence for substantial accretion around the WD in the binary plane of the system in the current stage.
The strong magnetic field of WD can dominate the accretion flow if associated magnetic energy density is comparable to the total kinetic energy density 
of the accreting plasma. This condition is specified by a characteristic distance called Alfv\'en radius which is expressed as \citep{Shapiro1983}  
\begin{eqnarray}
	r_A~&=&~\left(\frac{B_p^4 R_{wd}^{12}} {2GM_{wd}\dot m}\right)^{\frac{1}{7}} \nonumber\\
            &=&2\times10^9 \left(\frac{B_p}{10^8 \rm G}\right)^{\frac{4}{7}} \left(\frac{R_{wd}}{10^6 \rm cm}\right)^{\frac{12}{7}}
              \left(\frac{M_{wd}}{M_\odot}\right)^{-\frac{1}{7}} \left(\frac{\dot m}{10^{17} \rm g~s^{-1}}\right)^{-\frac{1}{7}}~\rm cm
\end{eqnarray}
where $B_p$ is the surface magnetic field of WD. In case of AR Scorpii, $r_A \sim 1.5\times10^{15}$ cm corresponding to $B_p = 7\times10^8$ G. 
This indicates that $r_A$ for magnetic WD in the binary system AR Scorpii is several orders of magnitude larger than the binary separation and 
radius of WD. The inner edge of accretion disk associated with WD in presence of strong magnetic field is placed at $r_A$. In side $r_A$, magnetic field 
dominated plasma moves along the magnetic lines of WD. The plasma frozen to magnetic field lines may sweep up the magnetic flux of WD leading to building up 
of magnetic stress far away from the stellar surface. This can ultimately halt the accretion flow on to WD from the companion star and accretion disk 
disappears at places where magnetic field controls the plasma. Therefore, WD in AR Scorpii may be currently in a propeller mass ejection phase with no 
mass transfer from the companion star. This together with higher value of $L_{sd}$ suggest that the WD binary system AR Scorpii is primarily rotation powered.

\subsection{Electromagnetic}
The gravitational properties of WD in AR Scorpii and its low X-ray luminosity suggest no mass loss or propeller mass ejection phase 
in the binary system. The non-thermal origin of X-ray emission from AR Scorpii \citep{Takata2018} and scenario of no mass transfer 
indicate a different physical process for draining the rotational energy from the rapidly spinning WD. This can be attributed to the 
magnetic dipole radiation from WD and magnetohydrodynamic (MHD) interactions \citep{Buckley2017}. The polar surface magnetic field 
of WD is given by \citep{Contopoulos2006}
\begin{eqnarray}
	B_p~&=&~\sqrt{\frac{3c^3M_{wd}P_s \dot P_s}{5\pi^2 R_{wd}^4}} \nonumber\\
	    &=&3.8\times10^{14} \left(\frac{M_{wd}}{M_\odot}\right)^{\frac{1}{2}}
	     \left(\frac{P_s}{117 \rm s}~\frac{\dot P_s}{3.8\times10^{-13} \rm s s^{-1}}\right)^{\frac{1}{2}}
	    \left(\frac{10^6 \rm cm}{R_{wd}}\right)^2 \rm G
\end{eqnarray}
under the assumption that the spin-down power is same as the energy loss due to a rotating magnetic dipole. 
The estimated value of surface magnetic field strength $B_p = 7\times10^8$ G using the observed parameters 
of AR Scorpii suggests that the WD in this binary system is highly magnetized. 
The estimated value of $B_p$ is broadly consistent with the upper limit ($\sim 5\times10^8$ G) derived 
from the polarimetric study of AR Scorpii by \citet{Buckley2017}. They suggest that variations present in the strong  
linear polarization measured from AR Scorpii can be described by a rotating magnetic dipole approximation of 
WD in the binary system. The bulk of spin-down luminosity is dissipated by the magnetic dipole radiation.
The strong magnetic field of the WD can generate the non-thermal high energy emission from AR Scorpii through the synchrotron 
radiation \citep{Singh2020}. The radius of WD light cylinder is defined as 
\begin{equation}
	R_{lc}~=~\frac{c P_s}{2\pi}~=~5.6\times10^{11} \left(\frac{P_s}{117 \rm s}\right)~ \rm cm
\end{equation}
For the spinning WD in AR Scorpii, $R_{lc} =5.6\times10^{11}$ cm is about eight times the binary or orbital 
separation of the system and implies that the companion star is well inside the magnetosphere of the WD. Therefore, 
the dense stellar wind produced by M-type companion star would have enough electric charge to screen out the inductive 
electric field unlike the isolated radio pulsars (rapidly rotating neutron stars) which produce very high electric potential 
drop along the magnetic field lines \citep{Fawley1977}. Under the dipole approximation of WD, the magnetic field at the light 
cylinder radius is given by
\begin{equation}\label{eqn:blc1}
	B_{lc}~=~\frac{\mu_{wd}}{R_{lc}^3}
\end{equation}	
where $\mu_{wd}$ is the magnetic dipole moment associated with the spinning WD. If WD spins down under magnetic dipole torques 
in a corotating plasma, $\mu_{wd}$ is described as \citep{Spitkovsky2006}
\begin{equation}
	\frac{1}{\Omega_s^3}\frac{d\Omega_s}{dt}~=~-\frac{\mu_{wd}^2}{c^3 I_{wd}}(1 + \rm sin^2 \alpha)
\end{equation}
The solution of this equation gives
\begin{equation}\label{eqn:dm}
	\mu_{wd}~=~\frac{1}{2\pi}\sqrt{\frac{c^3 I_{wd} P_s \dot P_s}{1 + \rm sin^2 \alpha}}
\end{equation}
where $\alpha$ is angle between magnetic and spin axes of the WD. For an observer at infinity, this oblique 
configuration exhibits a dynamic magnetic dipole moment which radiates its energy and the corresponding magnetic 
luminosity of WD is given by 
\begin{eqnarray}
	L_{mag}~&=& \frac{16 \pi^4 B_p^2 R_{wd}^6}{6 c^3 P_s^4} \nonumber\\
	       ~&=& 5\times10^{14}\left(\frac{R_{wd}}{10^6 \rm cm}\right)^6 \left(\frac{B_p}{10^8 \rm G}\right)^2
						  \left(\frac{117~ \rm s}{P_s}\right)^4 \rm erg~s^{-1}
\end{eqnarray}
For WD in AR Scorpii, we obtain $L_{mag} =2.8\times10^{33}$ erg~s$^{-1}$, which is same as the spin-down 
luminosity ($L_{sd}$) of the binary system. This suggests that the dipole approximation for WD in the binary system 
AR Scorpii can sufficiently provide the non-thermal luminosity measured from the source.
\par
From Equations (\ref{eqn:blc1}) and (\ref{eqn:dm}) we get
\begin{equation}\label{eqn:blc2}
	B_{lc}~=~4\pi^2 \sqrt{\frac{I_{wd} \dot P_s}{c^3 P_s^5(1 + \rm sin^2 \alpha)}}
\end{equation}	
Averaging over the inclination angle ($\alpha$), Equation (\ref{eqn:blc2}) can be written as 
\begin{eqnarray}
	B_{lc}~&=&~~4\pi^2 \sqrt{\frac{3 I_{wd} \dot P_s}{2 c^3 P_s^5}} \nonumber\\
	       &=&0.4\left(\frac{I_{wd}}{10^{50} \rm g~cm^2}\right)^{\frac{1}{2}} \left(\frac{117~ \rm s}{P_s}\right)^{\frac{5}{2}}
	       ~\left(\frac{\dot P_s}{3.8\times10^{-13}~ \rm s s^{-1}}\right)^{\frac{1}{2}}~\rm G
\end{eqnarray}
For the strongly magnetized WD in AR Scorpii, $B_{lc} \sim  0.7$ G is in good agreement with the dipole magnetic 
field of $\sim$ 0.4 at the light cylinder as estimated from the polarimetry \citep{Buckley2017}. This is very small as compared 
to the magnetic field at the light cylinder radius of neutron stars in the old as well as young radio pulsars.  
The corotation of plasma is not possible beyond $R_{lc}$ and therefore the magnetic field lines crossing the WD light cylinder open 
and converge to the polar caps on the surface of the WD. The opening angle associated with the polar caps of the WD is given as 
\citep{Ruderman1975}
\begin{equation}
	\theta_p~=~\sqrt{\frac{R_{wd}}{R_{lc}}}~=~0.08^\circ \left(\frac{R_{wd}}{10^6 \rm cm}\right)^{\frac{1}{2}}
	                                          \left(\frac{117~\rm s}{P_s}\right)^{\frac{1}{2}}
\end{equation}
and the radius of polar cap is defined as 
\begin{equation}
	R_p~=~R_{wd}\times \theta_p~=~3162 \left(\frac{R_{wd}}{10^6 \rm cm}\right)^{\frac{3}{2}}
	                              \left(\frac{10^{11} \rm cm}{R_{lc}}\right)^{\frac{1}{2}}~\rm cm
\end{equation}
In the binary system AR Scorpii, we get $\theta_p = 1.9^\circ$ and $R_p = 2.5\times10^7$ cm. 
The light cylinder and magnetic polar cap are high and low altitude sites respectively for the acceleration of particles to relativistic 
energies. The structure of field near the light cylinder is not known completely in radio pulsar however $B_{lc}$ is assumed to be very 
important for the $\gamma$-ray emission from these sources. The polar cap models for the high energy $\gamma$-ray emission from well known 
radio pulsars are based on the assumption that an accelerating gap exists above the magnetic polar cap in the open field line region. 
The region constrained by open field lines on the surface of WD (or neutron star) is referred to as polar cap area which is given by 
\begin{equation}
	A_p~=~2\pi R_{wd}^2 (1 - \rm cos \alpha)~=~6.28\times10^{12} \left(\frac{R_{wd}}{10^6 \rm cm}\right)^2(1 - \rm cos \alpha)~ \rm cm^2
\end{equation}	
The polar cap area estimated as $A_p = 3\times10^{16} \rm cm^2$ for magnetized WD in AR Scorpii is even less than 1$\%$ of the total 
surface area of the WD.  

\subsection{Other Exotic Properties}
The optical spectra of AR Scorpii show absorption lines varying sinusoidally in radial velocity on the orbital 
period ($P_o =$ 3.56 hours) of the binary system and the atomic emission lines in the UV/optical spectra originate 
from the companion M-type star \citep{Marsh2016}. WD is not visible in the absorption/emission spectra of AR Scorpii. 
The cosmological redshift of these spectral lines due to expansion of the Universe will be negligible since AR Scorpii 
is a Galactic source. However, the gravitational redshift ($z_g$) from the surface of WD for a distant observer is 
given by \citep{Ozel2013} 
\begin{equation}
	1 + z_g~=~(1-\eta_{wd})^{-\frac{1}{2}}
\end{equation}	
For the compactness of WD in AR Scorpii, the gravitational redshift $z_g =$ 0.00015, is very tiny. This suggests that 
wavelength of a photon emitted from the surface of WD will experience negligible shift with respect to the laboratory value. 
The spin frequency of WD ($\nu_s = P_s^{-1}$ = 8.54 mHz) represents high frequency component of the amplitude spectra of the 
pulsations at UV/optical/IR and radio wavebands, whereas the low frequency component is described by the beat frequency 
\begin{equation}
	\nu_b~=~\nu_s - \nu_o
\end{equation}	
where $\nu_o = P_o^{-1}$ is the orbital frequency of the binary system. The beat component with $\nu_b =$ 8.46 mHz is stronger 
than the WD spin component and dominates the pulsation period of AR Scorpii. The beat period represents orbital side band of the 
WD spin period. Red and blue-shifted hydrogen emission lines on timescales of WD spin/beat period are detected with radial 
velocities of $\sim$ 700 km $s^{-1}$ relative to the companion star \citep{Garnavich2019}. High temporal resolution spectroscopy 
suggests a very a complex line structure of the hydrogen emission lines observed from the binary system AR Scorpii. 
However, the physical mechanism that accelerates hydrogen to such high velocities is not clearly known. 
\par
\begin{figure}[t]
\centerline{\includegraphics[width=60mm,angle=-90]{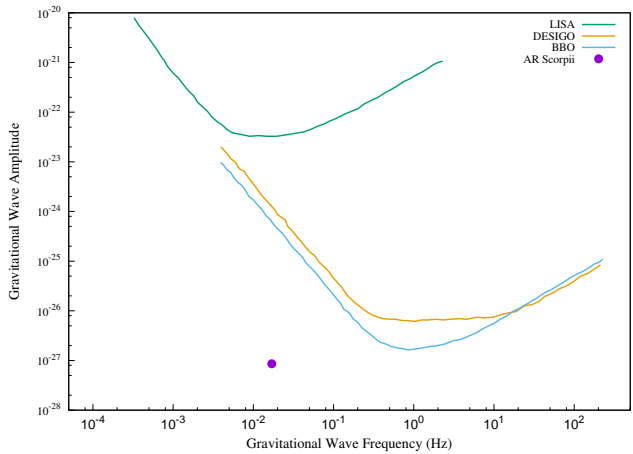}}
\caption{Dimensionless characteristic strain sensitivities curves for LISA \citep{Robson2019}, DECIGO/BBO \citep{Yagi2011} 
          correponding to a signal-to-noise ratio of 8 and integration time of 1 year along with the estimated gravitational wave amplitude 
          for AR Scorpii due to magnetic deformation of WD.\label{fig}}
\end{figure}

Strongly magnetized and rapidly spinning WDs find special interest in the gravitational wave astronomy. The intense 
magnetic field of WD causes asymmetry around its rotation axis. This asymmetry can generate gravitational waves due to 
the deformation of WD induced by the strong magnetic field in the binary systems like AR Scorpii. 
The amplitude of emitted monochromatic gravitational waves with a frequency $\nu_{GW} = 2 \nu_s$, is given by \citep{Shapiro1983,Maggiore2008}
\begin{equation}
      h_{GW}~=~\frac{28\pi^2}{3c^4}\frac{B_p^2 R_{wd}^6 \nu_s^2}{d~M_{wd}}
\end{equation}
For the rapidly rotating magnetic WD in AR Scorpii, we have $\nu_{GW}$ = 0.017 Hz and $h_{GW} \approx 8.5\times10^{-28}$. 
In Figure \ref{fig} , a comparison of $h_{GW}$ for AR Scorpii with the sensitivities curves for proposed space-based 
gravitational wave detectors LISA (Laser Interferometer Space Antenna), DECIGO (DECI-hertz Interferometer Gravitational 
wave Observatory), and BBO (Big Bang Observer) is presented. The sensitivity curves for different detectors in Figure \ref{fig} 
correspond to a signal-to-noise ratio of 8 and integration time of 1 year. We observe that the gravitational wave amplitude produced from the 
spinning WD in AR Scorpii due to magnetic deformation mechanism is much below the sensitivties of space-based detectors in 
the gravitational wave frequency range from 10$^{-4}$ Hz to 10 Hz. 
A surface magnetic field of above 10$^9$ G is required for WD in AR Scorpii to produce gravitational waves with amplitudes that can 
be detected by the upcoming space detector such as BBO \citep{Sousa2020}.  

\section{Summary}\label{sec4}
WD in the binary system AR Scorpii is the first and only strongly magnetized WD pulsating at UV-optical-IR and radio 
wavelengths discovered so far. The properties of WD in AR Scorpii suggest that it is very young and its spin is slowing down 
on a timescale of 10$^7$ years. The spin-down timescale is shorter than the characteristic age of WD whereas the 
cooling age exceeds the probable age of WD. The surface and core temperatures of 10$^4$ K and 10$^7$ K respectively show 
nice synergy with the observed features of AR Scorpii. Despite the strong gravitational and magnetic fields of WD, the 
accretion is not the main source of radiation from AR Scorpii, however accretion from a previous stage is necessary for 
spun up of WD because a WD is generally born with spin. The spinning WD is assumed to power most of the non-thermal emission 
from AR Scorpii. Under magnetic dipole approximation of WD, its spin-down luminosity can sufficiently drive the 
broadband emission from this source. The energy transfer from the fast spinning WD is either through the collimated 
out flow of relativistic particles (jet) or by the direct inductive interaction between the magnetospheres of WD and 
companion star. The emission originates from the most dissipation region near the 
companion M-star \citep{Geng2016,Katz2017,Takata2017,Garnavich2019,Singh2020}. 
Therefore, the spin of magnetic WD plays a very important role in the pulsed emission of AR Scorpii modulated over the spin/beat 
frequency. AR Scorpii remains a very promising and challenging source yet to be fully explored, however, various physical 
properties of WD investigated in this study would provide future guidance to explore the features of WD binary systems 
like AR Scorpii and AE Aquarii. These sources are also possible candidates for gravitational wave detection using better  
sensitive detectors in future. The broadband electromagnetic radiation together with the gravitational waves would better 
reveal the properties of compact objects in the binary systems.

\section*{Acknowledgments}
Author are thankful to the anonymous reviewers for their important and critical comments that greatly helped 
to improve the manuscript. 
\bibliography{MSR1}%
\end{document}